# Multi-wavelength Q-plate Arithmetic in an All-Liquid-Crystal Modular Setup

**JACEK PIŁKA, MICHAŁ KWAŚNY, MAGDALENA CZERNIEWICZ, MIROSŁAW KARPIERZ, AND URSZULA LAUDYN\***

*Faculty of Physics, Warsaw University of Technology, Koszykowa 75, 00-662 Warsaw, Poland*
*\*Corresponding author: urszula.laudyn@pw.edu.pl*

**ABSTRACT**

Vortex beams are a type of structured light characterized by phase rotation around the propagation axis, resulting in orbital angular momentum. Their properties make them useful in various applications such as high-resolution microscopy, optical tweezing, and telecommunications. This has led to a comprehensive development of methods for their generation, ranging from using single-purpose glass elements to utilizing computer-generated holograms using spatial light modulators. One of the most commonly used elements for vortex transformation is a vortex half-wave retarder called a *q*-plate, which can transform a Gaussian beam into a scalar vortex or vector beam depending on the input polarization. Although the commercially available ones are limited in the range of possible output topological charges, they can be stacked to perform arithmetic operations to expand them. However, changing the output or working wavelength requires rearranging the elements. We present an improvement to this method that solves these problems by introducing *Q*-modules, easy-to-fabricate, electrically tunable liquid crystal devices that combine the features of *q*-plates and half-wave plates and can be used as building blocks in modular assemblies. Electrical tuning makes it possible to change the working wavelength as well as the topological output charge or polarization order without the need to interact mechanically with the setup.

## Introduction

Light beams that have complex properties such as amplitude, phase, and polarization are currently of great interest in various fields. These beams have a wide range of applications, including optical communication systems [1], manipulating micro-objects [2], creating singular holograms, and efficient material processing [3].

Among other complex optical fields, vortex beams (VB) are structured light with a phase singularity in the center due to a phase rotation around the propagation axis [4]. This results in a characteristic doughnut-shaped intensity profile with a dark area in the center [4] and carrying an orbital angular momentum (OAM) [5]. An optical vortex is mainly characterized by its topological charge l, which is equal to the number of complete phase rotations at the distance of a single wavelength, and is proportional to the OAM it carries. The value of the VB charge also affects the beam's spatial profile. Although vortices are defined by their phase rotation, a specific class of them, called vector beams, are also attractive due to a spatially variant polarization rotation around their centers [6] that allows them to be focused to a narrower focal spot than the diffraction limit for a Gaussian beam (GB) of the same wavelength [7].

The specific shape and the presence of OAM gave the vortex and vector beams a wide range of applications, among which the most acknowledged are microscopy [8] and optical trapping [9], but also other fields such as cryptography [10]. This makes such beams a highly exploitable type of light, which in turn has led to the development of several methods of their generation. These methods can be divided into two groups: active and passive. The active ones consist of techniques involving spatial light modulators (SLMs) [11–13] and digital micro-mirrors devices (DMDs) [14] as dynamic holograms. Despite a somewhat extensive optical system configuration, both methods allow for precise manipulation of the output light. On the other end, passive elements such as spiral phase plates [15], static holograms [16,17], or axially-symmetric spatial half-wave plates called q-plates [18] are easy to integrate into various optical setups but do not provide the flexibility of active devices. The most straightforward applications of *q*-plates are to convert typical Gaussian beams into VB or to modify the OAM of transmitted vortex beams. In this aspect, *q*-plates are similar to conventional spiral phase plates but have several advantages, such as polarization control of the generated VB, electrical switching, and wavelength tuning capability. The passive *q*-plate elements typically consist of nematic liquid crystal (NLC) vector phase retarders [17], which are birefringent waveplates with a patterned distribution of the NLC optical axis. The NLC *q*-plate is essentially a liquid crystal cell with specially prepared alignment layers, fulfilled with an NLC. For structures characterized by an axially symmetric alignment of the molecules, such NLC *q*-plates act as spatially varying polarization converters that transform a Gaussian beam into a vortex or VB, according to the input polarization (circular or linear, respectively). The concept of *q*-plates originated from general considerations of spin angular momentum (SAM) and OAM exchange in inhomogeneous anisotropic media [19–21].

Nematic liquid crystals are fluids and anisotropic materials characterized by a non-random molecular spatial orientation described by the unit molecular vector, called director ***n*** [22]. While typical NLC cells provide a uniform director direction across the whole volume of the cell, a *q*-plate has a specific azimuthal distribution of ***n*** in the plane of the cell. The $\theta$ angle between ***n*** and a fixed reference axis (the *x*-axis) in the cell plane is the azimuthal coordinate, defined relative to a central point, the coordinate origin. As the NLC molecules are well anchored in the plane of the cell, the polar coordinate $\alpha = \frac{\pi}{2}$ remains constant for all the molecules. The local distribution of the director represents the local optical axis of the *q*-plate and can be derived from the formula ***n*** $= (sin\alpha cos\theta, sin\alpha sin\theta, cos\alpha)$.

The exact value of the $\theta$ angle in the plane of the *q*-plate can be determined from the transverse azimuthal coordinate $\phi = \arctan\left(\frac{y}{x}\right)$, and a linear function expresses $\theta$ and $\phi$ as follows: $\theta(\phi) = q\phi + \theta_0$, where the $\theta_0$ represents the orientation of the director on the *x*-axis, and the coefficient *q*, called the *q*-value, is the principal parameter of the *q*-plate. The *q*-value is an integer or half-integer and represents the number of revolutions of the director ***n*** in the $\theta$ angle range from 0 to $2\pi$. The sign of q indicates the direction of rotation. The effect of the *q*-plate on the plane wave is similar to that of the half-wave plate with a spatially variable optical axis. It can be represented as a matrix in Jones' formalism as follows [23]:

$$\mathbf{M}(q) = \begin{bmatrix} cos(2q\theta) & sin(2q\theta) \\ sin(2q\theta) & -cos(2q\theta) \end{bmatrix} \quad (1)$$

The NLC *q*-plate converts linearly polarized light into a vectorial beam, as shown in the following examples for a plane wave of horizontal and vertical polarization:

$$\begin{bmatrix} cos(2q\theta) & sin(2q\theta) \\ sin(2q\theta) & -cos(2q\theta) \end{bmatrix} \begin{bmatrix} 1 \\ 0 \end{bmatrix} = \begin{bmatrix} cos(2q\theta) \\ sin(2q\theta) \end{bmatrix} \quad (2)$$

$$\begin{bmatrix} cos(2q\theta) & sin(2q\theta) \\ sin(2q\theta) & -cos(2q\theta) \end{bmatrix} \begin{bmatrix} 0 \\ 1 \end{bmatrix} = \begin{bmatrix} sin(2q\theta) \\ -cos(2q\theta) \end{bmatrix} \quad (3)$$

On the other hand, if the input light is circularly polarized, the output field gets an additional $e^{\mp i2q\theta}$ component, indicating the rotation of the phase. An optical vortex is thus generated, characterized by a topological charge ±2q, where the sign depends on the handedness of the circular polarization, as shown in the example of the transformation of right-handed polarized light:

$$\begin{bmatrix} \cos(2q\theta) & \sin(2q\theta) \\ \sin(2q\theta) & -\cos(2q\theta) \end{bmatrix} \frac{1}{\sqrt{2}} \begin{bmatrix} 1 \\ -i \end{bmatrix} = \frac{1}{\sqrt{2}} \begin{bmatrix} e^{-i2q\theta} \\ ie^{-i2q\theta} \end{bmatrix} \quad (4)$$

The conversion of the input states considered above is also shown in Fig. 1Fig. 1. The linear polarization of the propagating light, schematically presented in the form of horizontal and vertical arrows in the leftmost column of Fig. 1Fig. 1(a,b), is transformed by the *q*-plate into an axially symmetric state. The transformation result can be identified by the additional analyzer at the output, as demonstrated on successive panels of Fig. 1(a,b) in the form of |4*q*| intensity maxima. Regardless of the *q*-value the phase of the beam remains unchanged.

The opposite behavior is observed when the input light is circularly polarized, as shown in Fig. 1Fig. 1(c). In this case, the output polarization remains unchanged, but the phase acquires a rotational component of a 2*q* frequency, causing the generation of the optical vortex with a *2q* topological charge.

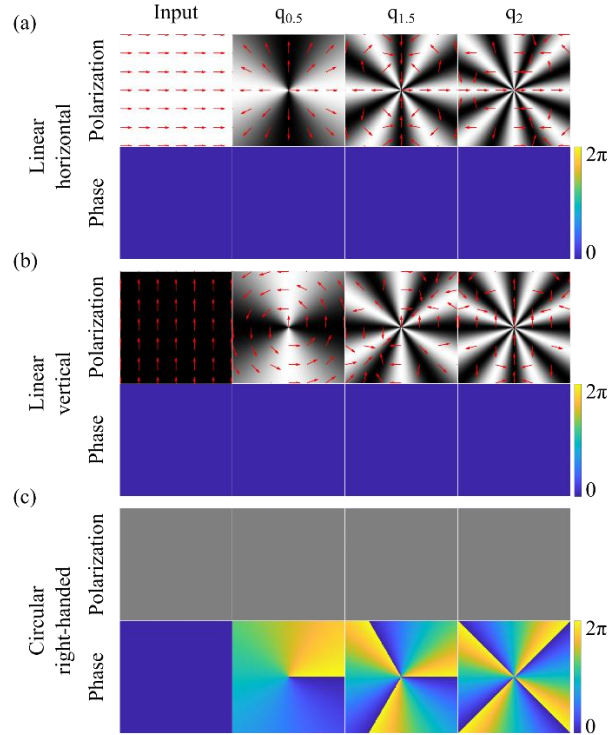

Fig. 1. Transformation of polarization and phase of a plane wave by *q*-plates of *q*-value 0.5, 1.5, and 2, shown for (a) a linear horizontal, (b) a linear vertical, and (c) a circular right-handed input polarization. The top/bottom panels in (a-c) represent the output light intensity profile after propagation through a horizontally oriented analyzer and the output phase, respectively. Red arrows indicate the electric field direction for the linear input polarization states in (a,b).

In typical commercial elements, the *q*-value is specified within the *q*-plate manufacturing process, meaning it is fixed. Changing the *q*-value would require a different *q*-plate. However, some optical setups' configurations may result in a change in the effective *q*-value. In one study, the *q*-value was doubled in a reflective geometry setup, where light passes through the *q*-plate twice, with an additional quarter-wave plate between the aforementioned phase plate and the mirror [24]. Another possibility is using a spatial light modulator (SLM) to dynamically encode *q*-plates at the desired *q*-value [25]. A disadvantage of this solution is the high cost of the SLM device and the subsequent requirement for a more complex optical system.

The commercially available *q*-plates are primarily available for low *q*-values of 0.5 and 1 and are designed to operate at wavelengths within a narrow spectral range. The first limitation can be overcome by implementing arithmetic operations on *q*-plates in the setup with additional half-wave plates [23]. A combination of two *q*-plates represented by *q*-values $q_a$ and $q_b$ with an additional half-wave plate in between would act as a single *q*-plate with a *q*-value equal to $q_a + q_b$:

$$\mathbf{M}(q_b) \cdot \mathbf{HWP} \cdot \mathbf{M}(q_a) = \begin{bmatrix} \cos(2q_b\theta) & \sin(2q_b\theta) \\ \sin(2q_b\theta) & -\cos(2q_b\theta) \end{bmatrix} \cdot \begin{bmatrix} 1 & 0 \\ 0 & -1 \end{bmatrix} \cdot \begin{bmatrix} \cos(2q_a\theta) & \sin(2q_a\theta) \\ \sin(2q_a\theta) & -\cos(2q_a\theta) \end{bmatrix}$$
$$= \begin{bmatrix} \cos(2(q_a + q_b)\theta) & \sin(2(q_a + q_b)\theta) \\ \sin(2(q_a + q_b)\theta) & -\cos(2(q_a + q_b)\theta) \end{bmatrix} = \mathbf{M}(q_a + q_b) \quad (5)$$

On the contrary, in a configuration where a half-wave plate follows two *q*-plates, the subtraction operation of *q*-values could be applied, which is given by the formula:

$$\mathbf{HWP} \cdot \mathbf{M}(q_b) \cdot \mathbf{M}(q_a) = \begin{bmatrix} 1 & 0 \\ 0 & -1 \end{bmatrix} \cdot \begin{bmatrix} \cos(2q_b\theta) & \sin(2q_b\theta) \\ \sin(2q_b\theta) & -\cos(2q_b\theta) \end{bmatrix} \cdot \begin{bmatrix} \cos(2q_a\theta) & \sin(2q_a\theta) \\ \sin(2q_a\theta) & -\cos(2q_a\theta) \end{bmatrix}$$
$$= \begin{bmatrix} \cos(2(q_a - q_b)\theta) & \sin(2(q_a - q_b)\theta) \\ \sin(2(q_a - q_b)\theta) & -\cos(2(q_a - q_b)\theta) \end{bmatrix} = \mathbf{M}(q_a - q_b) \quad (6)$$

The experimental implementation of the *q*-value addition and subtraction operations, according to equations (5) and (6), is schematically sketched in Fig. 2Fig. 2Fig. 2(a,b), respectively. Using the combinations of two *q*-plates and a half-wave plate, as shown in Fig. 2(a,b), a Gaussian beam can be transformed into a vector or vortex beam in a more comprehensive polarization order or topological charge range with a reduced number of different vortex retarders. Regardless of the method used to generate optical vortices, changing the beam's wavelength involves adjusting or replacing the phase elements to operate appropriately at a given optical frequency. To overcome this drawback, tunable phase elements, i.e., *q*-plates and half-wave plates, can be used, which can be easily adjusted to the desired wavelength. In a conventional configuration, as shown in Fig. 2, the transition between the addition and subtraction modes is realized by rearranging the order of the components, especially the position of a half-wave plate [23].

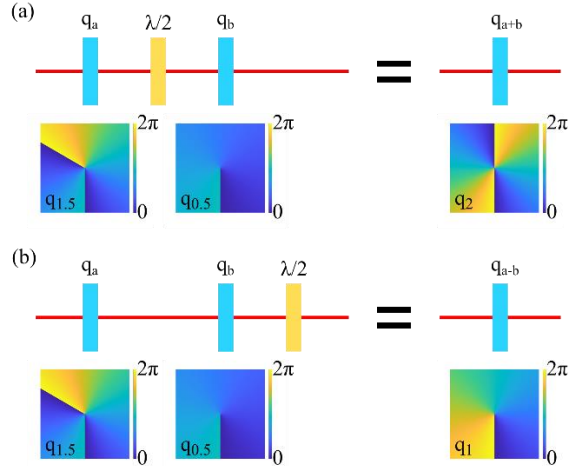

Fig. 2 The principle of arithmetic operations on *q*-plates in the configuration with an additional half-wave plate. Realization of *q*-values (a) addition and (b) subtraction operations. The bottom panels represent the fast axis distribution of utilized *q*-plates.

In our work, we further develop the existing q-plate arithmetic method that eliminates their two main drawbacks: the narrow working spectrum and the need for mechanical interaction when changing the operation type. To achieve this goal, we have replaced the classic vortex retarders and half-wave plates with electrically controlled NLC components called *Q*-modules. A single nematic liquid crystal *Q*-module contains two NLC cells: the first is a *q*-plate, and the second is a half-wave plate. We present a new concept for performing arithmetic operations on the resultant *q*-value using two or more *Q*-modules in a cascade arrangement. We show that by stacking *Q*-modules with different *q*-values, we can realize an optical device capable of performing mathematical operations such as addition and subtraction on vector vortex light beams. Our findings significantly contribute to optics and pave the way for new applications of liquid crystal *q*-plates in optical transmission systems, materials processing, atmospheric research, and more. This work demonstrates the potential for *Q*-modules to become essential components in advanced optical systems, leading to new developments in the field of optics and photonics.

## NLC Q-modules

The examples of proposed *Q*-modules are shown in Fig. 3(a,b). They are devices containing two NLC cells: a *q*-plate (left-hand cell) and a wave-plate (right-hand cell). The *q*-values of *q*-plates characterize the *Q*-modules. In the presented example, the *Q*-modules of $q = 1.5$ and $q = 0.5$ are shown, respectively.

For experiments, we prepared *Q*-modules using cells 12μm thick filled with 6CHBT NLC mixture [26]. The single NLC cell comprises two transparent glass plates coated inside with a transparent indium-tin-oxide (ITO) layer, which serves as electrodes. By using a slowly varying electric field (1kHz frequency), a controllable reorientation of the NLC molecules can be achieved, resulting in modification of the optical parameters of the cell. To achieve the required uniformity of the NLC cell thickness, dielectric micro-spacers with an average thickness of 12μm were inserted between the two glass substrates before gluing.

The spatially variant distribution of the local optical axis across the surface of the NLC *q*-plate was obtained by applying specially designed alignment layers realized with SD1 azo-dye (Dainippon Ink and Chemicals) and the photoalignment technique [27]. The SD1 molecules exhibit excellent adhesion to ITO film. When exposed to UV light, the azo-dye molecules reorient to a position perpendicular to the light polarization's direction. The very-low thickness of the alignment layer, of the order of a few nanometers, makes the optical retardation of the layer negligible. The utilized photoalignment technique [28] can write any pattern on the SD1-covered substrate as long as it is characterized by cylindrical symmetry; in particular, it allows for preparing *q*-plates with any integer or semi-integer, positive or negative *q*-value.

As a half-wave plate, we used an NLC cell characterized by a homogenous (planar) distribution of the molecules obtained by the unidirectional rubbing technique [29]. The element was designed to have an optical axis parallel to the zero-degree fast axis of the q-plate, as schematically presented in Fig. 3(a-b).

Two of the described cells, the *q*-plate and the wave-plate are then stacked to form an NLC *Q*-module. An external voltage can be applied to both parts, allowing independent reorientation of the LC molecules with respect to the propagation axis. This results in a change of their optical parameters, i.e., the phase delay with respect to the two orthogonal polarization components of the transmitted beam. This can be used either to tune the cell to operate within the desired wavelength or, if the voltage is high enough to achieve a homeotropic configuration (the Freedericksz threshold [22]), to turn it off completely. In this state, the elements induce neither a phase shift nor a polarization change, i.e. they are in the "off-state". Both states are shown on waveplates in Fig. 3(a,b). The first picture shows the waveplate in the "on-state" (molecules oriented neither in planar nor in homeotropic configuration). In the second, the element is in the "off state" (homeotropic orientation). In the cells tested, switching off was done by applying $7V_{rms}$.

Multiple *Q*-modules can be used together to create setups for arithmetic operations on q-values, similar to those shown in Fig. 2. This concept is illustrated in Fig. 3(c,d). However, because the half-wave plates can be turned off electrically, changing the type of operation (addition/subtraction) is done by applying high voltage to different elements without rearranging their order.

For the experimental verification of the concept, four *q-plate*s with $q = \pm 0.5, 1.5, 2$, and two NLC-plates were prepared, which were then combined to act as *Q*-modules in the experiments on arithmetic operations on *q*-values.

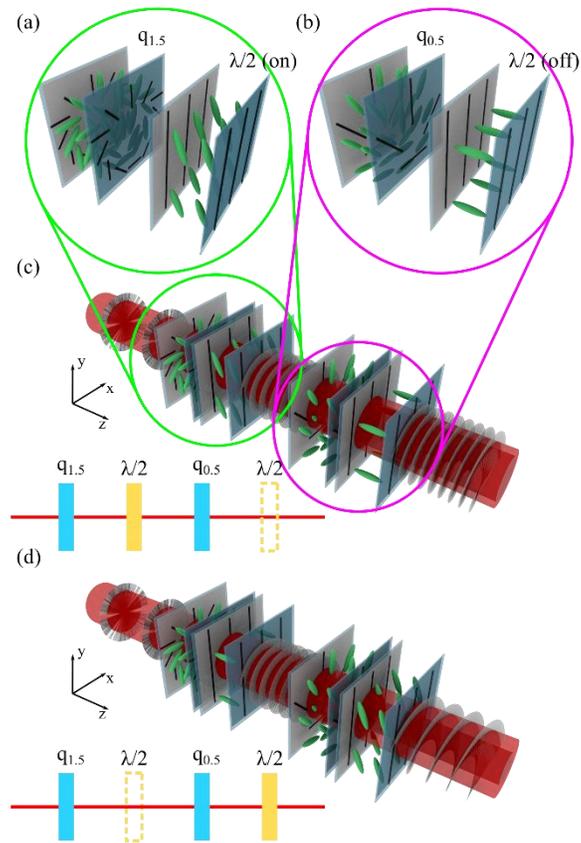

Fig. 3. (a,b) Schematic representation of the *Q*-modules, composed of two electrically controlled (on/off) NLC elements: a *q*-plate (left-hand-side) and a half-wave plate (right-hand cell). In-line combination of two LC-Q modules to perform (c) the addition and (d) the subtraction of *q*-values without re-positioning individual NLC cells. The corresponding configuration diagrams are shown in the bottom panels. The dashed yellow rectangle indicates the off-state of the half-wave plate.

First, the *q*-plates were tested to verify their correct operation. This was done in a typical Mach-Zehnder (MZ) interferometer with a *q*-plate inserted in one of the arms to observe the characteristic fork-like interference pattern of the output vortex beam. A CCD camera recorded the output field distribution to obtain both the intensity of the transformed VBs and their interference pattern. As the input wave, we used continuous wave Gaussian beams of two wavelengths: $\lambda = 1064$nm and $\lambda = 532$nm, which corresponds to the experimental results summarized in Fig. 4(a,b), respectively. An additional quarter-wave plate was used to switch the input polarization between circular and linear.

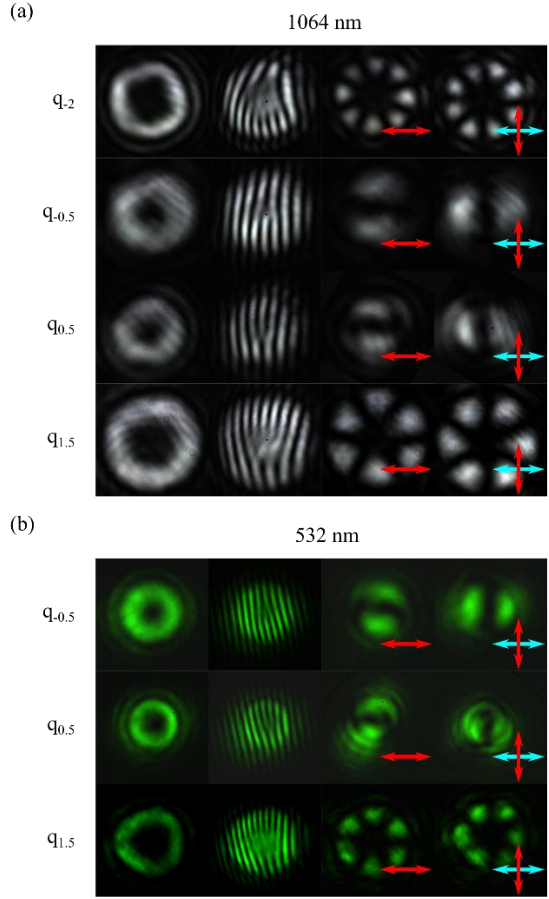

Fig. 4 Scalar and vector VBs generated with a single *Q*-module, obtained from a GB of a wavelength: (a) 1064nm and (b) 532nm. Each column shows (from the left): the intensity profile at the output VB generated from the right-handed circularly polarized light, the interference pattern of the VB with a linearly polarized Gaussian reference beam, the horizontal and vertical polarization component of the output vector VB beam generated from a linearly polarized GB. Polarization azimuths are marked by arrows: blue ones correspond to the input and red ones to the output analyzer.

An input circularly polarized Gaussian beam was transformed into a doughnut-shaped scalar VB (leftmost column, Fig. 4(a,b)), where the bright ring diameter increases with the *q*-value. An additional quarter-wave plate was used to convert the transformed VB to a beam with the same linear polarization as the reference to interpose a VB beam in the MZ interferometer, which confirms the designed value of a q-number. A rough analysis of the interference fringes shows a characteristic fork shape with 2*q* additional fringes (second column from the left, Fig. 4(a,b)), indicating a topological charge of 2*q*. The direction of the phase rotation, i.e., the sign of the topological charge, determines the direction of the bifurcation of the fringes. This can be seen, for example, in *Q*-modules with *q*-values of $q = -0.5$ and $q = 0.5$ (Fig. 4(a)), where an additional fringe appears at the bottom and top of the recorded image, respectively.

On the other hand, a linearly polarized Gaussian beam is transformed into a vector beam. To characterize a polarization distribution of the generated vector VBs, the images were taken with an additional analyzer placed between the *Q*-module and the CCD camera. Then, the recorded optical field profiles show a |4*q*| intensity maxima (two rightmost columns for the axis of the analyzer, Fig. 4(a,b)), indicating the correct operation of the prepared NLC components. The obtained results are consistent with equations 2 and 4.

The results for the visible light beam, shown in Fig. 4(b), are obtained for the same *Q*-modules as for the infrared light (Fig. 4(a)), which was made possible by tuning the NLC *q*-plates by the external electric field of the appropriate intensity. Thus, a single *Q*-module can operate over a wide spectral range within a visible and near-infrared band. The required AC voltage amplitude to effectively change the NLC cells' anisotropy remains low, within a range of $0 - 4$ V$_{rms}$ for the 6CHBT NLC mixture.

## Arithmetic operations using Q-modules

The experimental results of arithmetic operations using a set of *Q*-modules were investigated for two wavelengths, λ = 1064nm and λ = 532nm, and summarized in Fig. 5 and Fig. 6, respectively. Both of these figures are organized in the same way. The leftmost column shows the output intensity distribution of the generated VB resulting from adding or subtracting two q-values for an input circularly polarized Gaussian beam; the respective *q*-values are indicated on the left side in Fig. 5 and Fig. 6. The second left column shows the interference pattern of the VB with the Gaussian reference beam, with the appropriate additional fringes matching the order of an observed VB beam. The third and fourth columns represent the parallel and perpendicular linear polarization components of the output vector beam obtained from a linearly polarized input GB. Comparing the results presented in Fig. 5 and Fig. 6 with those recorded for a single *q*-plate (Fig. 4) it can be seen that using more NLC components does not negatively affect the quality of the generated beams. The output profiles, representing scalar and vector VBs of topological charge and polarization order, are equal to the sum/difference of the *q*-values of the individual NLC *Q*-modules.

Among the presented output VB field profiles, a special case may be of particular interest since it leads to a zero charge beam. It results from GB transformation through two *Q*-modules characterized by opposite q-value. In this case, a circularly polarized GB is transformed into a doughnut-shaped one, similar to an optical vortex with no specific phase rotation. This observation is confirmed by the absence of an interference fringe bifurcation, as shown in the third row of Fig. 5 and the second row of Fig. 6. For the linearly polarized input GB, the transformation preserves a linear beam polarization as seen at the output; however, the intensity profile is slightly distorted by the appearance of a darker circle at the center of the beam.

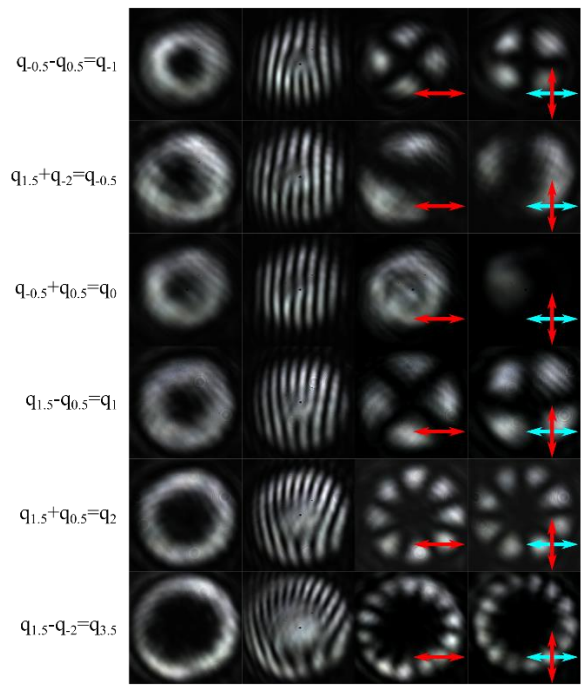

Fig. 5 The experimental results of examples of arithmetic operations using the *Q*-modules from Fig. 4 in configurations from Fig. 3(c-d), recorded for λ = 1064nm. Each column shows (from the left): the intensity profile at the output – i.e., VB, generated from the circularly polarized GB; the interference pattern of the VB with the reference GB; the horizontal and vertical polarization component of the output vector VB beam generated from a linear polarized GB. Polarization azimuths are marked by arrows: blue ones correspond to the input and red to the output axis direction of the analyzer.

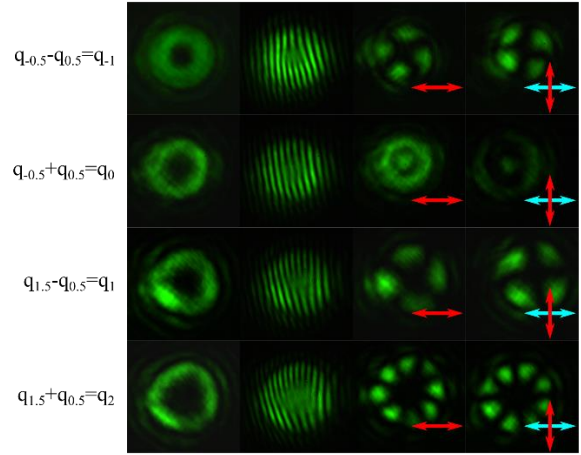

Fig. 6 The same as in Figure 5, recorded for λ = 532nm.

Although the tests were performed on operations with two NLC *Q*-modules, for clarity, not all possible combinations are shown in Figures 5 and 6. As shown in Table 1, such NLC components can be freely stacked in larger quantities, significantly increasing the number of achievable effective *q-value*s.

| *q*-values | Effective *q*-values |
|---|---|
| −0.5, 0.5 | −1, −0.5, 0, 0.5, 1 |
| −0.5, 0.5, 1.5 | −2.5, −2, −1.5, −1, −0.5, 0, 0.5, 1, 1.5, 2, 2.5 |
| −2, −0.5, 0.5, 1.5 | −4.5, −4, −3.5, −3, −2.5, −2, −1.5, −1, −0.5, 0, 0.5, 1, 1.5, 2, 2.5, 3, 3.5, 4, 4.5 |

Table 1 Possible *q*-values realized due to arithmetic with two different NLC *Q*-modules.

To demonstrate the scalability potential of the presented method, we performed a substraction operation with the use of three NLC *Q*-modules ($q = -2$, 0.5, and 1.5, respectively), as presented in Fig. 7(a). The general sequence of the operations that have been performed is illustrated in Fig. 2(b). The generated optical field results from a few-step transformation, first through the first two *Q*-modules characterized by *q*-values $q = -2$ and $q = 0.5$, followed by a third *Q*-module with *q*-value $q = -1.5$. The following panels (from left) of Fig. 7(b) show its intensity, phase, and polarization distribution. With any combination of only two prepared *Q*-modules, the charge −8 of the generated VB ($q = -4$) would not be achievable.

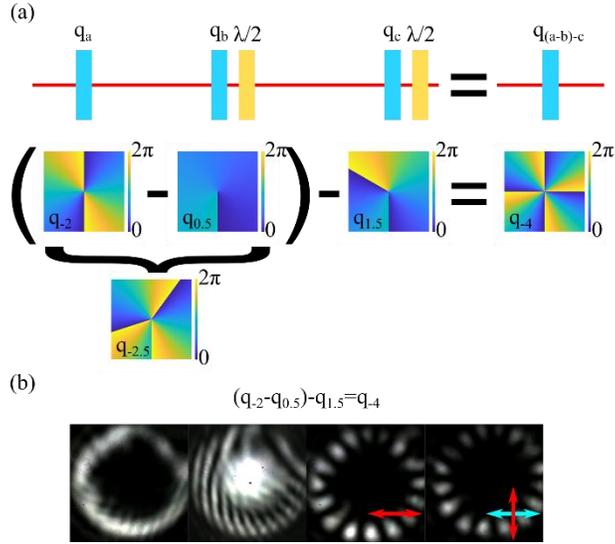

Fig. 7 Subtraction operation using three *q-plate*s. (a) A sketch of the sequence of *Q*-modules (top panel) and the fast axis distributions of the corresponding *q-plate*s of $q = -2$, $q = 0.5$, and $q = 1.5$ acting as single $q = -4$ q-plate (bottom panel). (b) From the left: the intensity profile of the generated VB generated from a circularly polarized GB, the interference pattern of the VB with a linearly polarized Gaussian reference beam, the horizontal and vertical polarization component of the output vector VB beam generated from a linearly polarized GB. Arrows indicate the polarization azimuths: blue arrows correspond to the input analyzer, and red arrows correspond to the output analyzer directions.

The scalability potential of the presented approach for switching between *q*-values of the VB is very high since the combined liquid crystal *q*-modules can perform subtraction, addition, and mixed combination and addition operations that the electric field can dynamically switch. As shown in Fig. 7, the combination of just three *Q*-modules characterized by $q = -2$, $q = 0.5$, and $q = 1.5$ supports up to eight different output *q*-values. Furthermore, by combining even more NLC Q-modules, this number increases, and obtaining any required output charge of the VB using a set of the elements of limited q-values becomes possible.

## Conclusions

In summary, we have demonstrated an arithmetic of *q*-values effectively realized in a wide spectral range (visible and near-infrared). To achieve this, we used a setup based on electrically controllable NLC modules that combine an NLC *q*-plate and a half-wave plate. Such *Q*-modules can be stacked like building blocks and act as a single *q*-plate with a resulting *q*-value according to the realized arithmetic operation. Thus, using a set of easily fabricated and well-operated NLC elements can be used for a vector and scalar VBs generation with a wide range of topological charges. Furthermore, unlike the originally presented method [23], by using electrical control, there is no need to rearrange the NLC optical components to switch between different charges of the generated VB at the output nor with the wavelength change.

Based on the use of vortex and vector beams, the presented method has the potential to be used in various laboratory applications. In addition, *q*-plate arithmetic is a promising area of research. It can potentially lead to new applications in quantum computing and other fields. By manipulating the polarization of light with *q*-plates, complex arithmetic operations can be performed in a spatially varying manner.

## CRediT authorship contribution statement.

**Jacek Piłka:** Methodology, Conceptualization, Data curation, Writing – original draft. **Michał Kwaśny:** Methodology, Investigation, Writing – original draft. **Magdalena Czerniewicz:** Investigation. **Mirosław Karpierz:** Supervision. **Urszula Laudyn:** Methodology, Conceptualization, Supervision, Funding acquisition, Project administration, Writing – review & editing.

## Funding.

This work was supported by the National Science Centre (grant agreement UMO-2016/22/M/ST2/00261) and by CB POB FOTECH 1 of Warsaw University of Technology within the Excellence Initiative: Research University (IDUB) program.

## Declaration of Competing Interest.

The authors declare that they have no known competing financial interests or personal relationships that could have appeared to influence the work reported in this paper.

## Data availability.

Data underlying the results presented in this paper are not publicly available at this time but may be obtained from the authors upon reasonable request.

## References

[1] T. Lei, M. Zhang, Y. Li, P. Jia, G.N. Liu, X. Xu, Z. Li, C. Min, J. Lin, C. Yu, H. Niu, X. Yuan, Massive individual orbital angular momentum channels for multiplexing enabled by Dammann gratings, Light Sci Appl. 4 (2015) e257–e257. https://doi.org/10.1038/lsa.2015.30.
[2] M. Woerdemann, C. Alpmann, M. Esseling, C. Denz, Advanced optical trapping by complex beam shaping, Laser & Photonics Reviews. 7 (2013) 839–854. https://doi.org/10.1002/lpor.201200058.
[3] A.L. Tolstik, Singular Dynamic Holography, Russ Phys J. 58 (2016) 1431–1440. https://doi.org/10.1007/s11182-016-0665-3.
[4] H. Rubinsztein-Dunlop, A. Forbes, M.V. Berry, M.R. Dennis, D.L. Andrews, M. Mansuripur, C. Denz, C. Alpmann, P. Banzer, T. Bauer, E. Karimi, L. Marrucci, M. Padgett, M. Ritsch-Marte, N.M. Litchinitser, N.P. Bigelow, C. Rosales-Guzmán, A. Belmonte, J.P. Torres, T.W. Neely, M. Baker, R. Gordon, A.B. Stilgoe, J. Romero, A.G. White, R. Fickler, A.E. Willner, G. Xie, B. McMorran, A.M. Weiner, Roadmap on structured light, J. Opt. 19 (2017) 013001. https://doi.org/10.1088/2040-8978/19/1/013001.
[5] A.M. Yao, M.J. Padgett, Orbital angular momentum: origins, behavior and applications, Adv. Opt. Photon., AOP. 3 (2011) 161–204. https://doi.org/10.1364/AOP.3.000161.
[6] Q. Zhan, Cylindrical vector beams: from mathematical concepts to applications, Adv. Opt. Photon., AOP. 1 (2009) 1–57. https://doi.org/10.1364/AOP.1.000001.
[7] S. Quabis, R. Dorn, M. Eberler, O. Glöckl, G. Leuchs, Focusing light to a tighter spot, Optics Communications. 179 (2000) 1–7. https://doi.org/10.1016/S0030-4018(99)00729-4.
[8] M. Ritsch-Marte, Orbital angular momentum light in microscopy, Philosophical Transactions of the Royal Society A: Mathematical, Physical and Engineering Sciences. 375 (2017) 20150437. https://doi.org/10.1098/rsta.2015.0437.
[9] Y. Yang, Y. Ren, M. Chen, Y. Arita, C. Rosales-Guzmán, Optical trapping with structured light: a review, AP. 3 (2021) 034001. https://doi.org/10.1117/1.AP.3.3.034001.
[10] A. Sit, F. Bouchard, R. Fickler, J. Gagnon-Bischoff, H. Larocque, K. Heshami, D. Elser, C. Peuntinger, K. Günthner, B. Heim, C. Marquardt, G. Leuchs, R.W. Boyd, E. Karimi, High-Dimensional Intra-City Quantum Cryptography with Structured Photons, Optica. 4 (2017) 1006. https://doi.org/10.1364/OPTICA.4.001006.
[11] C. Maurer, A. Jesacher, S. Fürhapter, S. Bernet, M. Ritsch-Marte, Tailoring of arbitrary optical vector beams, New J. Phys. 9 (2007) 78–78. https://doi.org/10.1088/1367-2630/9/3/078.
[12] S. Fu, T. Wang, C. Gao, Generating perfect polarization vortices through encoding liquid-crystal display devices, Appl Opt. 55 (2016) 6501–6505. https://doi.org/10.1364/AO.55.006501.
[13] Y. Zhou, X. Li, Y. Cai, Y. Zhang, S. Yan, M. Zhou, M. Li, B. Yao, Compact optical module to generate arbitrary vector vortex beams, Appl. Opt. 59 (2020) 8932. https://doi.org/10.1364/AO.401184.

[14] X.-B. Hu, C. Rosales-Guzmán, Generation and characterization of complex vector modes with digital micromirror devices: a tutorial, J. Opt. 24 (2022) 034001. https://doi.org/10.1088/2040-8986/ac4671.

[15] M.W. Beijersbergen, R.P.C. Coerwinkel, M. Kristensen, J.P. Woerdman, Helical-wavefront laser beams produced with a spiral phaseplate, Optics Communications. 112 (1994) 321–327. https://doi.org/10.1016/0030-4018(94)90638-6.

[16] P. Chen, B.-Y. Wei, W. Ji, S.-J. Ge, W. Hu, F. Xu, V. Chigrinov, Y.-Q. Lu, Arbitrary and reconfigurable optical vortex generation: a high-efficiency technique using director-varying liquid crystal fork gratings, Photon. Res. 3 (2015) 133. https://doi.org/10.1364/PRJ.3.000133.

[17] P. Chen, Y.-Q. Lu, W. Hu, Beam shaping via photopatterned liquid crystals, Liquid Crystals. 43 (2016) 2051–2061. https://doi.org/10.1080/02678292.2016.1191685.

[18] G. Machavariani, Y. Lumer, I. Moshe, A. Meir, S. Jackel, Spatially-variable retardation plate for efficient generation of radially- and azimuthally-polarized beams, Optics Communications. 281 (2008) 732–738. https://doi.org/10.1016/j.optcom.2007.10.088.

[19] L. Marrucci, C. Manzo, D. Paparo, Optical Spin-to-Orbital Angular Momentum Conversion in Inhomogeneous Anisotropic Media, Phys. Rev. Lett. 96 (2006) 163905. https://doi.org/10.1103/PhysRevLett.96.163905.

[20] L. Marrucci, C. Manzo, D. Paparo, Pancharatnam-Berry phase optical elements for wavefront shaping in the visible domain: switchable helical modes generation, Appl. Phys. Lett. 88 (2006) 221102. https://doi.org/10.1063/1.2207993.

[21] F. Bouchard, H. Mand, M. Mirhosseini, E. Karimi, R.W. Boyd, Achromatic orbital angular momentum generator, New J. Phys. 16 (2014) 123006. https://doi.org/10.1088/1367-2630/16/12/123006.

[22] I.-C. Khoo, Liquid crystals, Third edition, John Wiley & Sons, Inc, Hoboken, NJ, 2022.

[23] S. Delaney, M.M. Sánchez-López, I. Moreno, J.A. Davis, Arithmetic with q-plates, Appl. Opt., AO. 56 (2017) 596–600. https://doi.org/10.1364/AO.56.000596.

[24] M.M. Sánchez-López, J.A. Davis, N. Hashimoto, I. Moreno, E. Hurtado, K. Badham, A. Tanabe, S.W. Delaney, Performance of a q-plate tunable retarder in reflection for the switchable generation of both first- and second-order vector beams, Opt. Lett. 41 (2016) 13. https://doi.org/10.1364/OL.41.000013.

[25] I. Moreno, M.M. Sanchez-Lopez, K. Badham, J.A. Davis, D.M. Cottrell, Generation of integer and fractional vector beams with q-plates encoded onto a spatial light modulator, Opt. Lett. 41 (2016) 1305. https://doi.org/10.1364/OL.41.001305.

[26] R. Dabrowski, J. Dziaduszek, T. Szczuciński, Mesomorphic Characteristics of Some New Homologous Series with the Isothiocyanato Terminal Group, Molecular Crystals and Liquid Crystals. 124 (1985) 241–257. https://doi.org/10.1080/00268948508079480.

[27] V. Chigrinov, H.S. Kwok, H. Takada, H. Takatsu, Photo-aligning by azo-dyes: Physics and applications, Liquid Crystals Today. 14 (2005) 1–15. https://doi.org/10.1080/14645180600617908.

[28] S. Slussarenko, A. Murauski, T. Du, V. Chigrinov, L. Marrucci, E. Santamato, Tunable liquid crystal q-plates with arbitrary topological charge, Opt. Express. 19 (2011) 4085. https://doi.org/10.1364/OE.19.004085.

[29] J. Hoogboom, J.A.A.W. Elemans, A.E. Rowan, T.H.M. Rasing, R.J.M. Nolte, The development of self-assembled liquid crystal display alignment layers, Phil. Trans. R. Soc. A. 365 (2007) 1553–1576. https://doi.org/10.1098/rsta.2007.2031.